\documentclass[twocolumn,preprintnumbers,secnumarabic,amsmath,amssymb,nofootinbib,floatfix]{revtex4}
\usepackage{varioref,exscale,latexsym,amsmath,amssymb}
\usepackage{graphicx}

\usepackage{graphicx}
\usepackage{dcolumn}
\usepackage{bm}

\def\beq{\begin{equation}}

\def\eeq{\end{equation}}

\def\beqa{\begin{eqnarray}}

\def\eeqa{\end{eqnarray}}

\begin{document}

\title{{\bf Gauge federation as an alternative to unification}}

\medskip\
\author{John F. Donoghue}
\email[Email: ]{donoghue@physics.umass.edu}
\author{Preema Pais}
\email[Email: ]{ppais@physics.umass.edu}
\affiliation{Department of Physics,
University of Massachusetts\\
Amherst, MA  01003, USA}

\begin{abstract}
We motivate and explore the possibility that extra
$SU(N)$ gauge groups may exist independently of the
Standard Model groups, yet not be subgroups of
some grand unified group. We study the running of the
coupling constants as a potential evidence for a common
origin of all the gauge theories. Several different example
are displayed. Some of the multiple options
involve physics at the TeV scale.
\end{abstract}
\maketitle

\section{Introduction}

The Standard Model employs the gauge groups $U(1),~SU(2),~SU(3)$. At the most
simplistic level, this leads one to ask: why not $SU(4), ~SU(5)...$ etc? If
gauge theories associated with these groups were to exist independently, they
would likely be unobserved at present because they could become confining at much higher
energies. For example, if pure $SU(4)$ gauge theory
shared the same coupling strength as obtained in $SU(2)$ and $SU(3)$ at their intersection point
($4.7 \times 10^{17}$~GeV), $SU(4)$ would be confining at $4 \times 10^8$~GeV.
Higher $SU(N)$ groups would confine at yet higher
energies.

The usual approach to the higher order gauge groups is to attempt to embed
the Standard Model group into a larger Grand Unification group\cite{Georgi}. In this case,
$U(1)\otimes SU(2)\otimes SU(3)$ emerge as the unbroken subgroups of a single larger group.
Although there is not really an historical
precedent for gauge unification\footnote{The combining of electric and magnetic fields
was not a unification of two gauge theories, but rather the identification of
the correct U(1) gauge theory. Likewise, the electroweak unification is
really gauge mixing instead of gauge unification because of the two separate gauge
groups.}, it remains a very attractive idea. Most of the present explorations of physics
beyond the Standard Model are predicated on the unification paradigm.

However, alternatives are possible. For example, if higher groups
such as $SU(4)$ etc. are added sequentially
and independently of the Standard Model groups, the fermions of the new groups
may modify the running of the couplings in such a way that the couplings
converge on a common value at high energy. This could be the signal of a common
origin for all the gauge theories. In this case, there may be a fundamental
explanation for the set of $SU(N)$ gauge theories without having them all be combined
into a single unification group.

We explore this possibility in the present
paper. We will refer to the alternative as {\it gauge federation}.
A federation is an alliance of nearly autonomous self governing units. In the
present context the gauge theories themselves are autonomous and independent at
low energies. However, by hypothesis, they have a common origin and
share the same coupling strength at some high energy\footnote{As an alternative to GUTs, we are tempted to
call this common proposed underlying theory the "Federation of Independent Groups"
or FIGs (we thank Gene Golowich for the acronym) which has the
advantage of allowing us to refer to
the
resulting new particles as ``figments''.}. The running of the
couplings are also related because the fermions carry charge under more than one
gauge group.

A motivation for this possibility comes from idea of emergence.  For example,
Holgar Nielsen \cite{holgar1, holgar2} has proposed an intriguing rationale for why we
see gauge theories at low energies. If one has a complicated, maybe random,
fundamental theory at high energies with fluctuations of many types, the only
excitations that could be expected to propagate at large distances are those
which are protected, by a symmetry, from picking up a large mass scale. Gauge
symmetries require that the gauge bosons be massless, and so if there are competing
types of fluctuations, those associated with non-gauge theories would be expected
to share the scale of the fundamental theory while gauge degrees of freedom could
propagate and be active at low energy. There have been some partial
realizations of this
idea in condensed-matter-like systems\cite{emergence} with theories
in which the ground state has photonic excitations
although the original theory did not have photons as degrees of freedom. If such theories
were to generate the $U(1)\otimes SU(2)\otimes SU(3)$ groups, it would be natural to produce
higher gauge groups also.

Much like the grand unification paradigm, this idea does not generate a unique
theory as a result. There are many possible theories that differ by adjustable
assumptions. Our goal is to explore some of these possibilities. We will find
many options that are successful. Some are relatively simple extensions of the Standard Model.
Some are able to achieve federation at the Planck scale, and some can converge at
infinite coupling. There are some options that have the $SU(4)$ group becoming strong
at the TeV scale. We treat these in
separate sections below. However, there are also some general
features that we attempt to summarize in the conclusion.

\section{Running couplings and the $U(1)$ ambiguity}

We will be exploring the running of the coupling constants $g_1, g_2, g_3.... g_N$ of
the $U(1),~SU(2),~SU(3), ....SU(N)$ gauge theories, starting at the scale
$M_Z$ and continuing up to high energy. We will use the one loop beta functions \cite{running}
so that the running of the couplings are described by
\begin{equation}
\alpha_N^{-1}(\mu) = \alpha_N^{-1}(M_Z) + 8\pi b_N \ln \frac{\mu}{M_Z}
\end{equation}
with the constant $b_N$ defined by
\begin{equation}
\frac{d}{d~\ln\mu}g_N(\mu) = -b_N g_N^3
\end{equation}
The beta functions for $N\ge 2$ have the form
\begin{equation}
b_N = \frac{11N -2n_f}{48\pi^2}
\end{equation}
where $n_f$ is the number of fermions in the fundamental representation active
at the energy scale $\mu$. For U(1) the results depend on the hypercharge assignments,
\begin{equation}
b_1 = -\frac{1}{96\pi^2}\sum_i[Y_{Li}^2+Y_{Ri}^2]
\end{equation}
where $Y_{L,R}$ are the hypercharge assignments for left and right chiral fields, with
the covariant derivative
\begin{equation}
 D_\mu \psi_{L,R}= [\partial_\mu +i\frac{ g_1}{2} Y_{L,R} A_\mu ] \psi_{L,R}~~.
\end{equation}
The contribution of the Standard model fields is
\begin{equation}
b_1 = -\frac{5}{12\pi^2}
\end{equation}
For starting values, we use\cite{RPP}
\begin{eqnarray}
\alpha_1 (M_Z) &=& \frac{\alpha_{QED}(M_Z)}{\cos^2\theta_W(M_Z)} = 0.0106  \nonumber \\
\alpha_2 (M_Z) &=& \frac{\alpha_{QED}(M_Z)}{\sin^2\theta_W(M_Z)} = 0.0338 \nonumber \\
\alpha_3 (M_Z) &=&  0.118
\end{eqnarray}

\begin{figure}[!htb]
\begin{center}
\includegraphics[height=5cm]{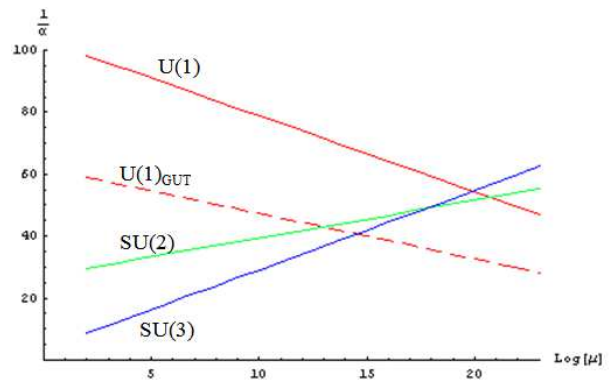}
\caption{The running of the inverse Standard Model coupling constants, with the GUT normalized $U(1)$ coupling also shown.}
\label{fig:couplings2}
\end{center}
\end{figure}

The normalization of the $SU(N)$ charges are well defined because the gauge bosons
carry the charge. However, for $U(1)$ this is not the case and we can choose any normalization that
we desire.  A rescaling of the coupling constant by a factor of $\rho$ would be accompanied by
a change in the hypercharge assignments by a factor of $1/\rho$. In the formulas above we have used
what we will call the ``Standard Model normalization'' corresponding to hypercharge assignments of $-1$ for the lepton
left handed doublets and $1/3$ for the quark left handed doublets. While this makes the $Z_0-\gamma$
mixing formulas look neat, there is really no
compulsion to use this normalization. For example, in describing the running coupling constants in
grand unified theories, most authors use the ``grand unification normalization'' convention,
\begin{equation}
g_1^{(GUT)} = \sqrt{\frac{5}{3}} g_1^{(SM)}
\end{equation}
which is appropriate for embedding the $U(1)$ group within the larger GUT group.
The running of the Standard Model charges in both normalizations is shown if Fig. 1.
However, this normalization
need not be appropriate for our efforts either.

This feature makes it clear that there is then an inherent ambiguity in our program, associated with
the normalization of the $U(1)$ charge. From low energy information only, we have no way of
knowing what the appropriate $U(1)$ charge normalization is\footnote{For amusement we note that if we rescale the $U(1)$
charge by a factor of $g_1^{(special)} = \sqrt{\frac{6}{5}} g_1^{(SM)}$ we would bring the running couplings
of the SM groups into reasonable concordance at $\mu = 4.7 \times 10^{17}$~GeV}. Because our goal is an exploration of the
various possibilities for the convergence of the couplings, we will allow ourselves to rescale the $U(1)$ charge
by integer ratios at times in this work. This freedom clearly opens up yet more possibilities than found in the present work. For the explorations of the present paper, we use the Standard Model normalization, as we feel that this is sufficient to illustrate the range of possible features.

\section{Simple extensions}

In this section we consider the simplest extensions of the gauge groups, where one continues to add higher
order groups. The examples cited will lead to a convergence of the coupling constants. Throughout the paper, we
do not insist on absolutely perfect convergence. Besides the fact that we use the leading order beta functions,
there likely would be threshold effects that modify the running near the federation point. Our convergence criteria is relatively
simple - if the running couplings converge within the thickness of the lines in our plots (which for $SU(3)$ is smaller
than the experimental error bars) we will accept the result as sufficiently converged.

\subsection{Sequential extensions}

\begin{figure}[!htb]
\begin{center}
\includegraphics[height=5cm]{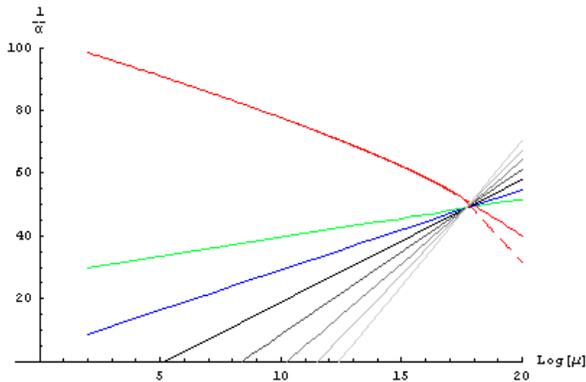}
\caption{The evolution of coupling constants with higher order groups up to $SU(90)$ (solid red line) and $SU(11000)$ (dashed red line) given $U(1)$ hypercharge. Higher order coupling constants are shown in grayscale.}
\label{fig:U1couplings1}
\end{center}
\end{figure}

In a first instance, we give the fermions of the higher groups a $U(1)$ charge only. Since the $SU(2)$ and $SU(3)$ running is
not modified, the convergence point for the couplings has to be the crossing of these theories in
the SM running $M_F= 4.7 \times 10^{17}$~GeV. For this case we chose to include 3 generations of fermions in each higher order
 $SU(N)$ group, with each generation containing one fermion with vector hypercharge coupling of  $+1/N$
and one of hypercharge $-1/N$. This situation readily has
no anomalies. The $1/N$ factors were tried because the quarks of $SU(3)$ carry hypercharge of multiples of $1/3$, and was
accepted because the resulting pattern led to gauge federations. We have included a ridiculously high number of
gauge groups - up to $SU(11000)$. The $U(1)$ coupling does meet up with the other
couplings at the convergence point, as can be seen in Fig \ref{fig:U1couplings1}. The high number of gauge groups is somewhat illusory because
comparable results can be obtained including a much smaller number, such as N=90. The high N groups run so fast
that they decouple almost immediately. Of more interest in the lightest gauge groups. The first few of these are also
shown in Fig. \ref{fig:U1couplings1}. $SU(4)$ becomes strong at the scale $\Lambda_4= 10^5$~GeV.

 In the above exercise, we have assumed that the fermion masses are of the same order as $\Lambda_N$. This is reasonable since
 the fermions are added in vector representations and there is no gauge symmetry that forces the masses to vanish, as happens
 in the Standard Model. The only natural
 scale in the theory is then $\Lambda_N$, and the fermion masses should be comparable to this scale.
 If this were not to be the case, there could in principle be light bosonic bound states that would influence the running
 below the scale of  $\Lambda_N$. We have also been careful to avoid anomalies in the gauge currents, as we will continue to
 do in what follows. Clearly the lack of gauge anomalies is another key restriction on possible quantum numbers of new fermions.

\begin{figure}[!htb]
\begin{center}
\includegraphics[height=5cm]{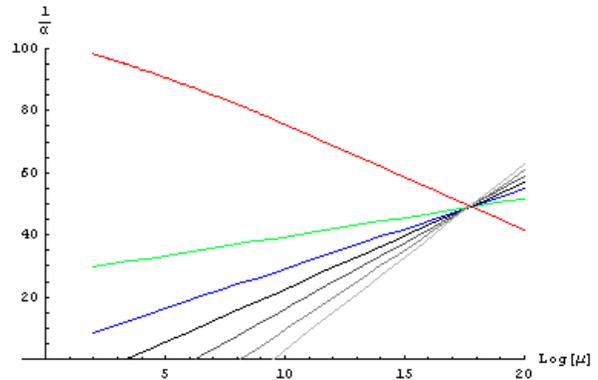}
\caption{The evolution of coupling constants with higher order groups up to $SU(7)$ with $N$ generations of vector fermions
and a hypercharge assignment of $\pm 1/N$.}
\label{fig:U1couplings2}
\end{center}
\end{figure}

 A second example of the same form is the same as the above example but includes N generations of fermions instead of 3 generations.
 Again the hypercharge assignments within a generation are $\pm 1/N$. Because the running is faster, we are not able to include
 as many gauge groups. We find that the convergence of the couplings occurs when we include groups up to N=7. The running  of the
 couplings is shown in Fig. \ref{fig:U1couplings2}. We see that  $SU(4)$ becomes strong at $\Lambda_4 = 2.5 $~TeV.
 The largest scale in the theory is then $\Lambda_9 =3 \times 10^9$~GeV.

\begin{figure}[!htb]
\begin{center}
\includegraphics[height=5cm]{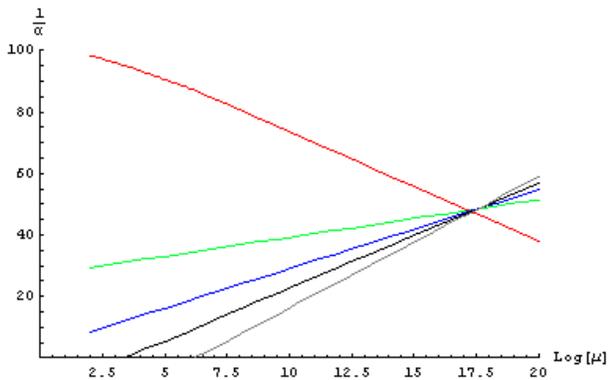}
\caption{Coupling constants with extra order gauge groups $SU(4)$ and $SU(5)$  with N generations and hypercharge $\pm 1/3$.}
\label{fig:U1couplings3}
\end{center}
\end{figure}

 For completeness, a related variation would have N generations but using hypercharges $\pm 1/3$ in each generation. In this case
 we get convergence with only groups up to N=5 with $\Lambda_4 = 2.5$~TeV and $\Lambda_5 = 1.7 \times 10^6$~GeV. The results are shown in
 Fig. \ref{fig:U1couplings3}.

\subsection{Prime $SU(N)$ couplings}

 Instead of simply extending the gauge groups sequentially to higher $N$, Nielsen and Brene\cite{holgar2} argue for a more specific
 pattern based on considerations on random dynamics. They make the case that the allowed groups must correspond to $N$ being
 a prime number. The Standard Model satisfies this. However, Nielsen and Brene
 are not able to argue that the series should stop at $N=3$. Therefore we should expect further $SU(N)$ groups with $N$ equal to a prime number.
 We will explore this case also.

\begin{figure}[!htb]
\begin{center}
\includegraphics[height=5cm]{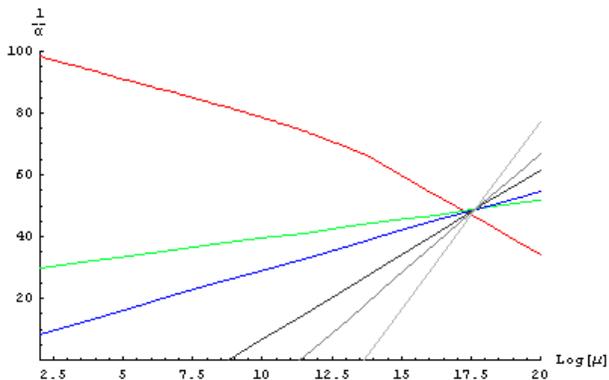}
\caption{Coupling constants with higher order gauge groups $SU(5),~SU(7)$ and $SU(11)$ with N vector fermions given a $U(1)$ hypercharge of $1/3$.}
\label{fig:primeU12}
\end{center}
\end{figure}

\begin{figure}[!htb]
\begin{center}
\includegraphics[height=5cm]{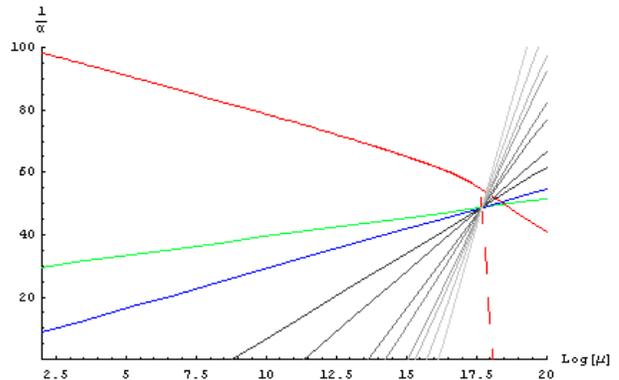}
\caption{The evolution of coupling constants with higher order groups up to $SU(113)$ (solid red line) and $SU(7919)$ (dashed red line) given $U(1)$ hypercharge. Higher order coupling constants are shown in grayscale}
\label{fig:primeU11}
\end{center}
\end{figure}

 In this picture, the next gauge group would not be $SU(4)$ but $SU(5)$. In Fig \ref{fig:primeU12} we show an example of such a theory
 with higher groups consisting of prime $N$ up to $N=11$. This uses the $N$ fermions with hypercharge $1/3$. If we change the hypercharge assignment
 to $1/N$, we can include groups up to extremely high values - probably infinite within the uncertainties. In
 Fig. \ref{fig:primeU11} we show the evolution of the couplings including groups up to $SU(7919)$ added.

\section{Planck scale federation}

An attractive possibility is that federation occurs at the Planck scale. In emergent theories, this would be
plausible if all the interactions, including the gravity and underlying space-time structure, were
emergent from a common underlying theory. There is added theoretical motivation for this option from the
Weinberg-Witten theorem\cite{Weinbergwitten}, which says that composite Yang-Mills gauge bosons or gravitons cannot be
formed from a Lorentz invariant theory. A neat way around this is if the space--time itself is emergent\cite{spacetime}.
The Planck scale would then be the dominant indicator of the underlying scale. The running of the gravitational
strength is dominantly quadratic in the energy because of the dimensionality of the coupling constant. New particles
will influence this modestly through a renormalization of Newton's constant, but will not change the dominant quadratic
running.
\begin{figure}[!htb]
\begin{center}
\includegraphics[height=5cm]{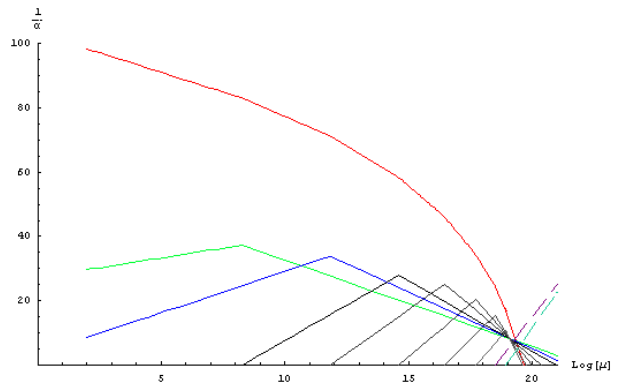}
\caption{The inverse coupling constants, with higher order $SU(N)$ added(up to $SU(10)$, each with $N$ vector fermions).}
\label{fig:planck1}
\end{center}
\end{figure}

In order to achieve convergence of the coupling at the Planck scale, we must also modify the running of
$SU(2)$ and $SU(3)$ by having some of the new fermions couple to these groups also. One pattern that
works is to allow the $SU(N)$ fermions to also carry a charge under $SU(N-2)$ In Fig \ref{fig:planck1} we illustrate one
solution that employs groups up to N=10 with there are N fermions of each hypercharge. Note the loss of asymptotic freedom that occurs because the fermions in the higher $SU(N)$ theory overwhelm the gauge contributions in the $SU(N-2)$ theory, such that the latter starts off asymptotically free and then changes when the $SU(N)$ contributions turn on.

Further examples where the federation point is the Planck scale are found in the next section.

\section{Infinite coupling}

 As one adds more gauge groups with more fermions, one can readily lose the property of asymptotic
 freedom. Some of the solutions shown above display this property. A coupling constant in a non-asymptotically-free
theory would eventually run to infinity (i.e. $1/\alpha_i \to 0$) if the running continues to high enough scale. This raises
the possibility that {\em all } the couplings could run to infinity at the federation point. Then the
low energy theories would emerge from a more primitive theory corresponding to infinite coupling. This would potentially
be an interesting option for the emergence idea sketched in the introduction. We give an example
of such a situation in this section.

There is a obvious numerical issue about the approach to strong coupling. We are
using the lowest order beta function. As the coupling gets strong, one needs higher order contributions
to the beta function. Approaching infinite coupling would require all orders, including non-perturbative
contributions. Clearly this full description is beyond our power. However, the one loop running can still be used
as a predictor of the energy scale when the theory enters strong coupling. There is then some inherent ambiguity
into the subsequent evolution, including the scale at which the coupling goes to infinity. This ambiguity
will be different for different gauge groups, which then leads to a certain fuzziness in the concept of
federation at infinite coupling. Given these limitations, we will present an example of a situation where
the lowest order running leads to unification at infinite coupling, and expect that there could be modest
deviations from this picture that would work in a more complete analysis.

\begin{figure}[!htb]
\begin{center}
\includegraphics[height=5cm]{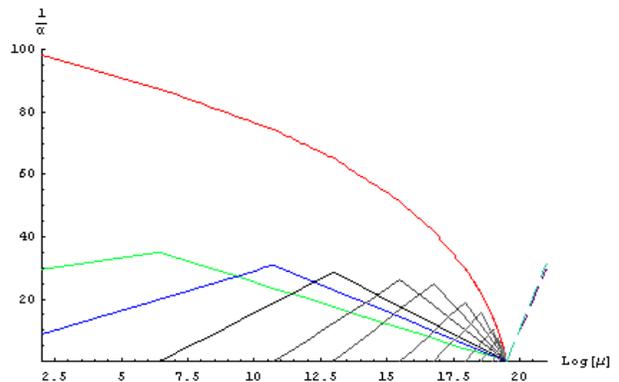}
\caption{Higher order $SU(N)$ groups added, with $N$ fermions, and $SU(N+2)$ fermions have $SU(N)$ charge, with the couplings meeting at Planck scale}
\label{fig:infcouplings2}
\end{center}
\end{figure}

Our first example has another nice feature - the couplings run to infinity at the Planck scale. In order to accomplish
this we introduce the new fermions of the $SU(N)$ group such that they also carry charges under
$SU(N-2)$ as well as $U(1)$. In this instance, we look at a model with $N$ vector fermions, with the meeting scale at around the Planck scale ($3.2\times10^{19}$). Groups up to $SU(19)$ were added, with $U(1)$ hypercharge $4/7$. The results are in Fig. \ref{fig:infcouplings2}

\begin{figure}[!htb]
\begin{center}
\includegraphics[height=5cm]{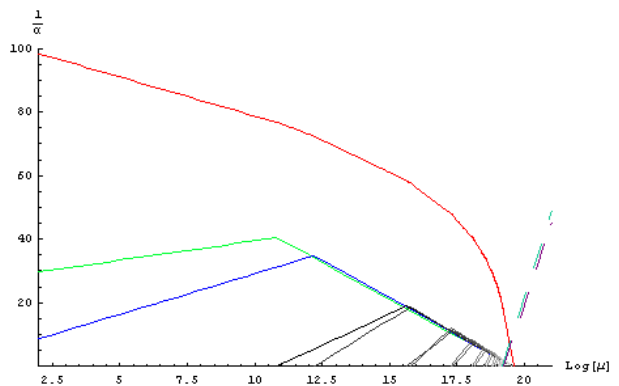}
\caption{Higher order $SU(N)$ groups have 6 fermions, with the couplings meeting at the Planck scale.}
\label{fig:infcouplings3}
\end{center}
\end{figure}

A similar example has a less unusual hypercharge assignment. In this next case we take 3 fermions with hypercharge +1/2 and three with -1/2, and include gauge groups up to $N=21$.
The resulting running couplings are shown in Fig
\ref{fig:infcouplings3}. The merging of the final group is
less accurate than we has found in other cases throughout the paper, but will accept it as illustrative of this possibility. The federation point is $3.3 \times 10^{19}$~GeV, close to the Planck scale.

\begin{figure}[!htb]
\begin{center}
\includegraphics[height=5cm]{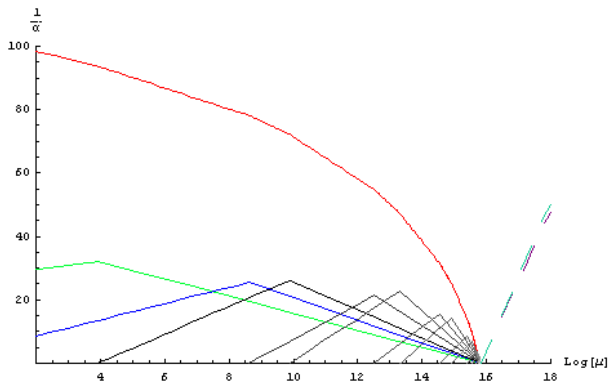}
\caption{Higher order $SU(N)$ groups added, with $N$ fermions, and $SU(N+2)$ fermions have $SU(N)$ charge. The $SU(4)$ cutoff is at TeV scale.}
\label{fig:infcouplings1}
\end{center}
\end{figure}

Another working option has a smaller federation scale. Again we give
the  $SU(N)$ fermions charges under
$SU(N-2)$, and include up to $N=21$, but include N fermions each of hypercharge $\pm 2/3$. The greater number of fermions leads
to a faster running and the coupling constants all reach infinite strength at $6.8 \times 10^{15}$~GeV, as seen in Fig. \ref{fig:infcouplings1}.
The $SU(4)$ coupling blows up at 8.4 TeV.

\section{SU(4) at the TeV scale}
The last figure of the previous section, Fig. \ref{fig:infcouplings1}, has a further interesting property - the $SU(4)$ coupling also
becomes infinite at $\Lambda_4 =8.4$ TeV. This situation can be found in other simulations also. For example both the cases of Fig \ref{fig:U1couplings2} and Fig \ref{fig:U1couplings3} have a mass scale of $\Lambda_4 =2.5$~TeV. We need not take these specific numbers too seriously, but it proved to be surprisingly common to find TeV scales for the $SU(4)$ group.

The phenomenology of this $SU(4)$ group is reasonably similar to that of earlier versions of Technicolor\cite{technicolor}. In these forms of Technicolor, there was also a new confining group at the TeV scale, with $SU(4)$ being a typical example. However, the technifermions were endowed with a chiral symmetry and were given chiral couplings to the electroweak group. In this case, the dynamical breaking of the chiral symmetry also lead to electroweak symmetry breaking. In our situation, we have used vectorial assignments which allows for a bare mass term for any $SU(4)$ fermion. The interaction of these fermions would not lead to symmetry breaking. However, the fermions are still coupled vectorially to either the hypercharge gauge boson or to the $SU(2)$ gauge bosons. While their production properties would depend on the specifics of the model, these couplings would lead to the expectation of producing the new fermions through $W,~Z,~\gamma$ interactions. In the cases that we have studied, the new $SU(4)$ fermions do not couple to gluons.

The spectrum of the fermions would be expected to be QCD-like. If the light quarks had a bare mass of order $\Lambda_{\rm QCD}$, quark model arguments indicate that the pseudoscalar states would still be the lightest mesons, although they would not be pseudo-Goldstone bosons as in QCD. The $J^{PC}=1^{--}$ vector state would be the next lightest, followed by the orbitally excited states. Because they can be produced directly from a vector
gauge boson, the vector bound states would be seen as a resonance in Drell-Yan production. Rates and signatures for this would be similar to those predicted for the Techni-rho states\cite{techniphenomenology}. These vector states are searched for in the $WZ\to \ell^+\ell^-\ell^\pm\nu,~Z\gamma\to \ell^+\ell^-\gamma, ~ZZ,~{\rm and}~ \mu^+\mu^-$ final states. The pseudoscalar states would decay into two gauge bosons\footnote{Because of our choice of vectorial couplings, the analog of of the axial-vector transition $\pi \to W^*\to e\nu$ is not available.}. In this case, the decay $P^0\to Z^0Z^0-$ would be the most visible.

Detailed studies of signals of dynamical symmetry breaking models show that many of the new hadronic states should be able to be uncovered at
the LHC \cite{techniphenomenology}. The specific details on individual channels depend on the detector properties. In general signals for dynamical symmetry breaking are somewhat uncertain because there is no compelling model with unique predictions. The couplings of our $SU(4) $ theory is similar to QCD-like Technicolor theories, up to modifications of order a factor of two because of our use of vector couplings rather than chiral couplings. These studies lead us to conclude that within the federation scheme if Nature places the $SU(4)$ scale in the scale of a few TeV, the LHC should be well suited to uncover evidence of this new physics.

\section{Conclusions}

We have shown that the running couplings can reach a common value under the influence of higher $SU(N)$ groups. These higher gauge groups would not have been seen yet because they decouple at higher energy. We have called this possibility gauge federation and argued that it may be an indicator of common underlying dynamics. It makes the most sense in the
context of emergent theories rather than unified theories. Without a principle to explain
why only the Standard Model gauge groups are emergent, we would also look forward to other
gauge groups.

We have found that it is
relatively straightforward to combine independent gauge theories in ways that do lead to a common coupling
at a high energy. While admittedly some of the successful combinations that we have found appear somewhat
random,
we conclude that there are many plausible ways to implement the idea of gauge federation. This
is both good and bad. It is unfortunate that there is not a very restricted possibility to achieve federation,
because if there were only a few instances there would be firmer predictions. However, on the
positive side it also means that it is more plausible that a fundamental theory could have this property,
and once uncovered, could lead to predictions that differentiate it from other theories.

In all cases however, there is a clear prediction of a hierarchy in the scales of the gauge interactions. The next lightest group is always $SU(4)$, which occurs at scales between a few TeV and $10^8$~GeV depending on the federation point and the choice of fermion content. At the lightest scales we might discover the particles directly. It would be interesting to explore whether the higher $SU(N)$ groups could also have useful implications, for example through modifications that could influence baryogenesis or inflation.

\section*{Acknowledgements} This work has been supported in part by the NSF grant PHY- 055304 and in part by the Foundational Questions Institute. We thank Stephane Willocq for information on the experimental tests of dynamical symmetry breaking. We also thank Andreas Ross and Koushik Dutta for discussions throughout this project.

\end{document}